\documentclass[sigconf]{acmart}

\usepackage{array}
\usepackage[english]{babel}
\usepackage[utf8]{inputenc}
\usepackage{blindtext}
\usepackage{caption}
\usepackage{subcaption}
\usepackage{multirow}

\usepackage{booktabs}

\usepackage{cleveref}

\usepackage{xspace}
\usepackage{graphicx}
\usepackage{balance}
\usepackage{layouts}

\let\myRef\ref
\renewcommand\ref{\unskip~\myRef}
\let\myCite\cite
\renewcommand\cite{\unskip~\myCite}

\newcommand{\ie}{i.e., \@}

\newcommand{\parax}[1]{\noindent \textbf{#1}}



\copyrightyear{2022}
\acmYear{2022}
\setcopyright{rightsretained}
\acmConference[IMC '22]{Proceedings of the 22nd ACM Internet Measurement Conference}{October 25--27, 2022}{Nice, France}
\acmBooktitle{Proceedings of the 22nd ACM Internet Measurement Conference (IMC '22), October 25--27, 2022, Nice, France}
\acmDOI{10.1145/3517745.3561452}
\acmISBN{978-1-4503-9259-4/22/10}

\begin{document}

\title{Illuminating Large-Scale IPv6 Scanning in the Internet}

\author{Philipp Richter}
    \affiliation{
        \institution{Akamai}
        \country{}
    }
    \email{prichter@akamai.com}
\author{Oliver Gasser}
    \affiliation{
        \institution{Max Planck Institute for Informatics}
        \country{}
    }
    \email{oliver.gasser@mpi-inf.mpg.de}
\author{Arthur Berger}
    \affiliation{
        \institution{Akamai/MIT}
        \country{}
    }
    \email{arthur@akamai.com}

\begin{abstract}
	
	While scans of the IPv4 space are ubiquitous, today little is known about scanning activity in the IPv6 Internet. In this work, we present a longitudinal and detailed empirical study on large-scale IPv6 scanning behavior in the Internet, based on firewall logs captured at some 230,000 hosts of a major Content Distribution Network (CDN). We develop methods to identify IPv6 scans, assess current and past levels of IPv6 scanning activity, and study dominant characteristics of scans, including scanner origins, targeted services, and insights on how scanners find target IPv6 addresses. Where possible, we compare our findings to what can be assessed from publicly available traces. 
	Our work identifies and highlights new challenges to detect scanning activity in the IPv6 Internet, and uncovers that today's scans of the IPv6 space show widely different characteristics when compared to the more well-known IPv4 scans.

\end{abstract}

\maketitle

\section{Introduction}

\label{sec:introduction}

Scanning the address space for vulnerable hosts and services is a key component in many of today's cyberattacks. In the IPv4 space, a scan of the entire address space can be conducted with comparably little resources in less than one hour~\cite{ZMap}, and botnets constantly scan the IPv4 space randomly to find new targets for infection~\cite{mirai-botnet-usenix-security}. This ubiquity of scanning activity in the IPv4 space makes scan detection readily possible, e.g., by leveraging darknets, or monitoring traffic on hosts or honeypots~\cite{richter2019scanning}. In the IPv6 Internet, both carrying out scans, as well as their detection, present a vastly more complicated task. Scanners can not simply target random addresses (there are more than $10^{38}$ IPv6 addresses) and must hence rely on hitlists or other heuristics to generate targets. At the same time, also the detection of IPv6 scans is challenging for two reasons: firstly, we need a vantage point that attracts and sees significant amounts of scanning traffic. 
Secondly, the vastness of the IPv6 space allows scanners to use entire subnets of varying sizes to emit scan traffic, potentially scanning from trillions of different source IP addresses, masking the true source of the scan traffic, and making scan detection difficult.
Thus, conflating IPv6 and IPv4 scans, while tempting, presents a false equivalence.
In this paper, we present a first-of-its-kind broad and longitudinal study of large-scale IPv6 scanning in the Internet. We make two key contributions: 

\parax{Illuminating IPv6 scanning activity:} We present detailed analyses on large-scale IPv6 scans carried out over the course of 15 months, as seen from a major CDN. We analyze scan sources, and study targeted services and addresses. We find that, unlike IPv4 scans, large-scale IPv6 scans are still comparably rare events, and we find them originating only from some 60 ASes. Further, IPv6 scan packets are concentrated on a small number of very active scan sources, with the two most active sources accounting for more than 70\% of all logged scan traffic throughout our measurement window. Many large-scale IPv6 scans do not target a single or a small number of specific services, but rather scan large swaths of port numbers, sometimes exceeding 100 ports targeted per scan. This behavior more closely resembles general and unspecific penetration testing behavior, as opposed to scanning patterns of botnets trying to spread laterally by exploiting individual vulnerabilities. Our initial findings show that IPv6 scans in the wild show widely different characteristics from the more well-known IPv4 scans. We contrast our findings with what can be observed in publicly available data, and discuss potential reasons for our observations.

\parax{Measurement methodology:} We identify key methodological challenges when it comes to pinpointing IPv6 scan sources and quantifying scanning activity and its properties. 
Regular IPv6 traffic is exchanged between two hosts using their 128-bit IPv6 addresses. However, in the case of scan traffic, we commonly find scanning actors not sourcing scan packets from an individual 128-bit source address, but from myriad source addresses spread across large prefixes. In such cases, any individual 128-bit source address used by a scanner may only emit very few packets (or even just a single packet), and thus hardly meet any criterion to be classified as a scan source.
In fact, we find scanners using source addresses spread across prefixes as unspecific as a /32 prefix, a typical IPv6 allocation size for an \textit{entire ISP}, thereby masking the true source of scanning activity.
We show that when not aggregating source addresses to less-specific prefixes, such scanning activity may be missed in part or entirely, and can lead to severe misinterpretation of findings. Yet, in turn, too coarse aggregation of sources leads to conflating individual scan actors as well as non-scanning hosts. 
The methodological challenges faced in this work directly apply to scan detection and blocking in operational settings (e.g., Intrusion Detection Systems) and we argue that they present a looming major challenge which could become a widespread problem if and when  vulnerability scanning in the IPv6 space becomes more common.

Our work constitutes an early view into IPv6 scanning, with much future work to be done, both academically, as well as operationally when it comes to hardening and securing systems in the IPv6 Internet.
The remainder of this paper is structured as follows: In Section~\ref{sec:detect} we introduce our datasets and detection mechanisms. We characterize IPv6 scans and their properties  ``in the wild'' in Section~\ref{sec:characteristics} and provide a comparison with what can be observed in publicly available data in Section~\ref{sec:mawi-crosscheck}. We conclude with a discussion of our findings, related work, and future work implications in Section~\ref{sec:discussion}.

\section{Detecting IPv6 Scanning}
\label{sec:detect}

In this section, we introduce our vantage point and datasets, pre-filtering, our large-scale scan definition, and a first-order overview of scanning activity.

\subsection{Vantage Point and Dataset}
\label{sec:vantage_point_and_dataset}

\noindent\textbf{CDN firewall:} 
Our primary data source are unsolicited IPv6 packets collected at the firewall of a subset of the machines of a major CDN. We collect any unsolicited incoming packets destined to port numbers other than TCP/80 and TCP/443 (since the machines support services on these port numbers).  Note that we also do not collect ICMPv6 packets.
Our data ranges from January 1, 2021 to March 15, 2022 and covers traffic logged at some
230,000 machines in over 700 ASes.\footnote{Note that over the course of our measurement window the number of
servers and ASes in the CDN change.  The considered subset of the $\approx$ 230,000 machines corresponds roughly to two thirds of the deployed servers and in half of the networks.} Each machine has IPv6 addresses assigned that are \textit{client-facing}, i.e., returned in DNS responses to clients accessing content at the CDN, as well as \textit{non client-facing} IPv6 addresses, which are never returned in DNS responses.

\begin{figure}
	\centering
	\includegraphics[width=0.99\linewidth]{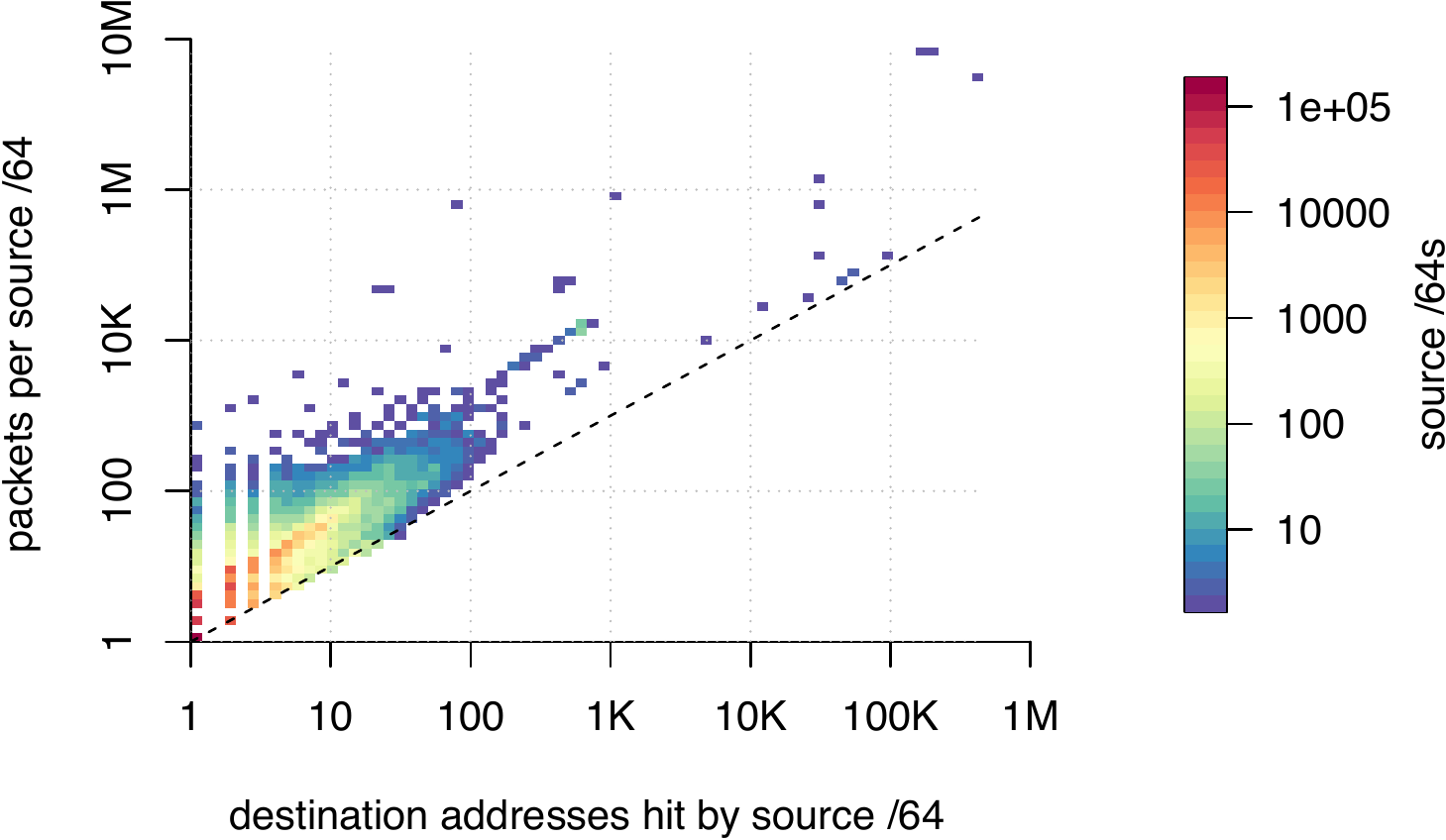}
	\caption{Heatmap of source /64s: number of destination IP addresses targeted (\textit{x}-axis) and number of packets logged (\textit{y}-axis) in November 2021.}
	\label{fig:srcstats}
\end{figure}

\parax{CDN artifact removal:} 
The exposed \textit{client-facing}
addresses are prone to \textit{(a)} connection artifacts by misconfigured Web clients\footnote{For example Web clients trying to establish IPsec connections or attempting NetBIOS name resolution with every outgoing TCP connection. See previous work~\cite{richter2019scanning} for an analysis of such phenomena in the IPv4 space.} and \textit{(b)} repeated failing connection attempts where clients or servers unsuccessfully try to access a service on a specific domain name (e.g., deliver email to a domain that does not accept email). The former, \textit{(a)} are removed by our requirement for sources to contact a large number of destination IPs to be considered a scan, see next section. To remove the latter, \textit{(b)}, we remove IPv6 /64 sources for which we log more than 30\% of ``5-duplicate'' packets, i.e., packets that hit the same destination IP and port more than 5 times over the course of a day. The two most common artifacts are connection attempts on ports TCP/25 (SMTP) and UDP/500 (IPsec). We opt for this port agnostic filtering method (i.e., not filtering packets based on port numbers), since any port number may also be targeted as part of actual scanning campaigns. We provide more details on filtering artifacts in~\Cref{sec:artifacts}.

\parax{Per-source statistics:} %
Figure~\ref{fig:srcstats} shows, for the month of November 2021, a histogram of all source /64s in our firewall logs, partitioned by the number of targeted destination IP addresses (\textit{x}-axis), and the total number of packets logged for a given /64 (\textit{y}-axis). Generally, we would expect that a source conducting a scan targets a large number of destination IP addresses. We can make two observations \textit{(i)} the majority of source /64s cluster close to the origin, targeting a very small number of destinations, and are unlikely to carry out scans. Rather, we can expect that many of the packets from these sources resemble aforementioned artifacts. \textit{(ii)} We only see a comparably small number of /64 sources that target a large number of destinations (see blue dots towards the right).

\subsection{Scan Detection}

\noindent\textbf{Scan definition:} In this work, we focus on large-scale scanning events, thus we define a \textit{scan} as a source targeting at least 100 destination IPv6 addresses with a timeout, or maximum packet inter-arrival time, of 3,600 seconds (1 hour). Here, we use a similar threshold to previous work in IPv4~\cite{richter2019scanning}.  We point out that other works have used less strict definitions of a scan, e.g., requiring a source to target fewer addresses, such as 25~\cite{darknet-imc15} or only 5~\cite{fukuda2018knocks}. We acknowledge that our strict large-scale definition may miss small-scale scans, but at the same time greatly reduces the number of CDN connection artifacts that we otherwise may mis-classify as scanning activity. We study the sensitivity of our scan detection parameters in the section below.

\noindent\textbf{Scan source aggregation:} A key aspect of defining an IPv6 scan source is the level of source aggregation, i.e., whether to treat each IPv6 source address (a /128) independently, or whether to aggregate all packets from an individual source prefix, say a /64 prefix, and then apply our scan detection. Scanners can use a single IPv6 address to source their packets from, or, depending on the address space available to them, send their scan packets from a multitude of different source addresses in a prefix, be it to encode specific scan information inside the source IP address, or to evade detection. For example, a scan actor may opt to select a random source IP address for each of its scan packets so that no single source address would show up, e.g., as a scan source in firewall logs of targeted networks.
Throughout this paper, we show statistics for detected scans when treating sources as individual /128 addresses, for /64 source aggregation, as well as /48 source aggregation (the smallest Internet routable entity in IPv6), and separately point out cases where none of these aggregations fit. Note that we first aggregate all packets for a given prefix, and subsequently apply our scan detection method, including our requirement for traffic to target at least 100 destination IP addresses. Thus, we can have situations in which a /48 prefix gets classified as scan source, while none of its contained /64s qualifies as a scan source, individually. The number of detected source /48s can, and does, hence exceed the number of source /64s. We point out that there is no ``one-size-fits-all'', and both a too-specific as well as a too-coarse aggregation carries risks of mis-identification and mis-attribution. 

\begin{table}%

	\begin{tabular}{r|r|r|r|r}
		\textbf{aggregation} & \textbf{scans} & \textbf{packets} & \textbf{sources} & \textbf{ASes} \\
		\hline
		/128 & 65,485 & 2.04B & 3,542 & 55 \\
		/64 & 5,199 & 2.14B & 1,326 & 62 \\
		/48 & 5,019 & 2.15B & 1,372 & 76 \\
		\hline
	\end{tabular}

	\caption{Detected scans over the course of our measurement window (Jan 2021 until Mar 2022). Depending on the aggregation of source IP addresses, the number of scans and scan sources changes dramatically.}
	\label{tab:scans}
\end{table}

\noindent\textbf{Scan totals:} Table~\ref{tab:scans} shows the number of scans, associated packets, sources, and scan source ASes when treating scan sources as /128, /64, or /48 prefixes. 
We can see that the level of aggregation has a major impact on the number of detected scans, as well as associated sources and networks. As a first-order observation, we
find that large-scale scanning campaigns hitting the CDN are much rarer compared to scans of the IPv4 space, e.g., previous work~\cite{richter2019scanning} detected \textit{in a single month} in 2018 more than 1 million IPv4 scan source addresses in total, out of which more than 1,000 sources were probing the entire IPv4 space.

\noindent\textbf{Parameter sensitivity:} To study the effect of our timeout threshold (1 hour), we compute the scan detection for /64 prefixes using shorter timeouts of 1,800 seconds (30 mins) as well as 900 seconds (15 mins). We find that using the 1,800 second threshold, we detect 5,175 scans (down 0.5\%) from 1,221 sources (down 8\%). For the 900 second threshold, we detect 5,097 scans (down 2\%) from 1,182 sources (down 11 \%). We hence note that the timeout settings have a comparably small impact on the number of detected scans. When relaxing our threshold of 100 destination IP addresses from 100 to 50, we detect 22,701 scan events (up 436\%) from 7,835 /64 sources (up 590\%), vastly more than with our threshold of 100 destination IPs. Closer inspection reveals that 7,210 (92\%) of these sources belong to a single AS (AS~\#18), which we study in more detail in Section~\ref{sec:scan_sources}.

\begin{figure}
	\centering
	\includegraphics[width=0.98\linewidth]{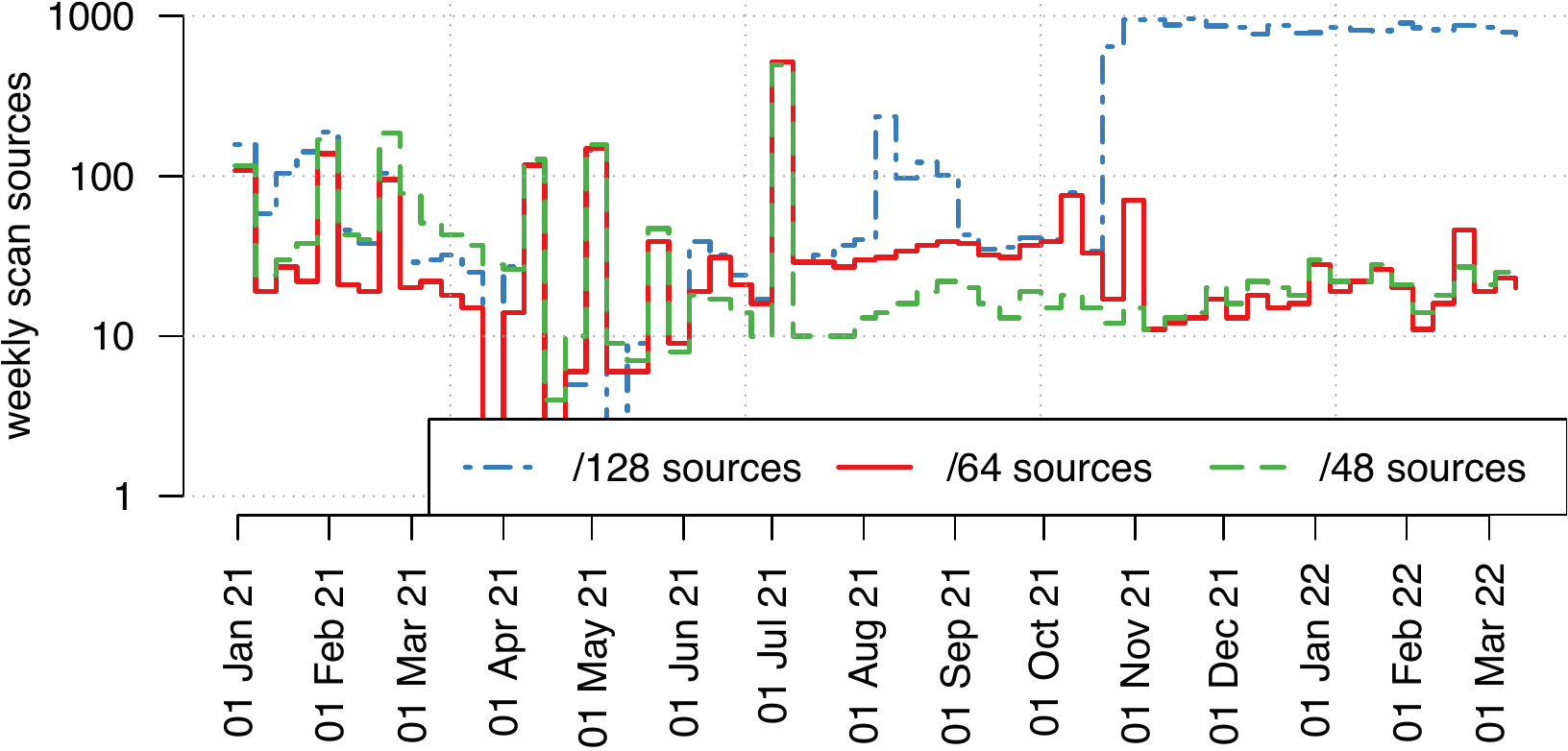}
	\caption{Weekly scan sources when first aggregating the traffic source by /128, /64, and /48 prefix, and then applying our scan detection.}
	\label{fig:weeklyscanners}
\end{figure}

\section{IPv6 Scan Characteristics}
\label{sec:characteristics}

Leveraging our longitudinal dataset, we first study scans over time, then we examine the sources of scans and assess which ports and addresses IPv6 scanners target. 

\subsection{Scans over Time}

\noindent\textbf{Long-term trend:} Figure~\ref{fig:weeklyscanners} shows the number of active IPv6 scan sources per week. In this figure, we aggregate all packets from source addresses by the respective /128, /64, and /48 prefix, and then apply our scan detection mechanism over the resulting set of packets from a given source. Overall, when aggregating source addresses by /64 or /48 prefix, we note that over the course of our measurement window we see a relatively constant number of active scan sources in the 10-100 range (median weekly active /64 sources is 22).
In terms of active /128 scan sources, we see a strong uptick starting in November 2021, which could naively be interpreted as an overall uptick in scanning entities.
However, we show later that this uptick is due to a single scan entity, see discussion of AS~\#9 of Table~\ref{tab:sourceases}. 
Note that depending on the aggregation level, the number of weekly active sources varies by almost two orders of magnitude. 

\noindent\textbf{Scan durations:} Without source aggregation, IPv6 scans are dominated by short scans (median duration 94 seconds), but the longest scan lasts for more than 128 days. This scanner is also one of the most active scanners in terms of packets, as we will discuss in the next section. When aggregating sources to /64 prefixes, the median scan duration increases to 2.7 hours, when aggregating to /48 prefixes, the median duration increases to 3.4 hours.

\begin{figure}
	\centering
	\includegraphics[width=0.97\linewidth]{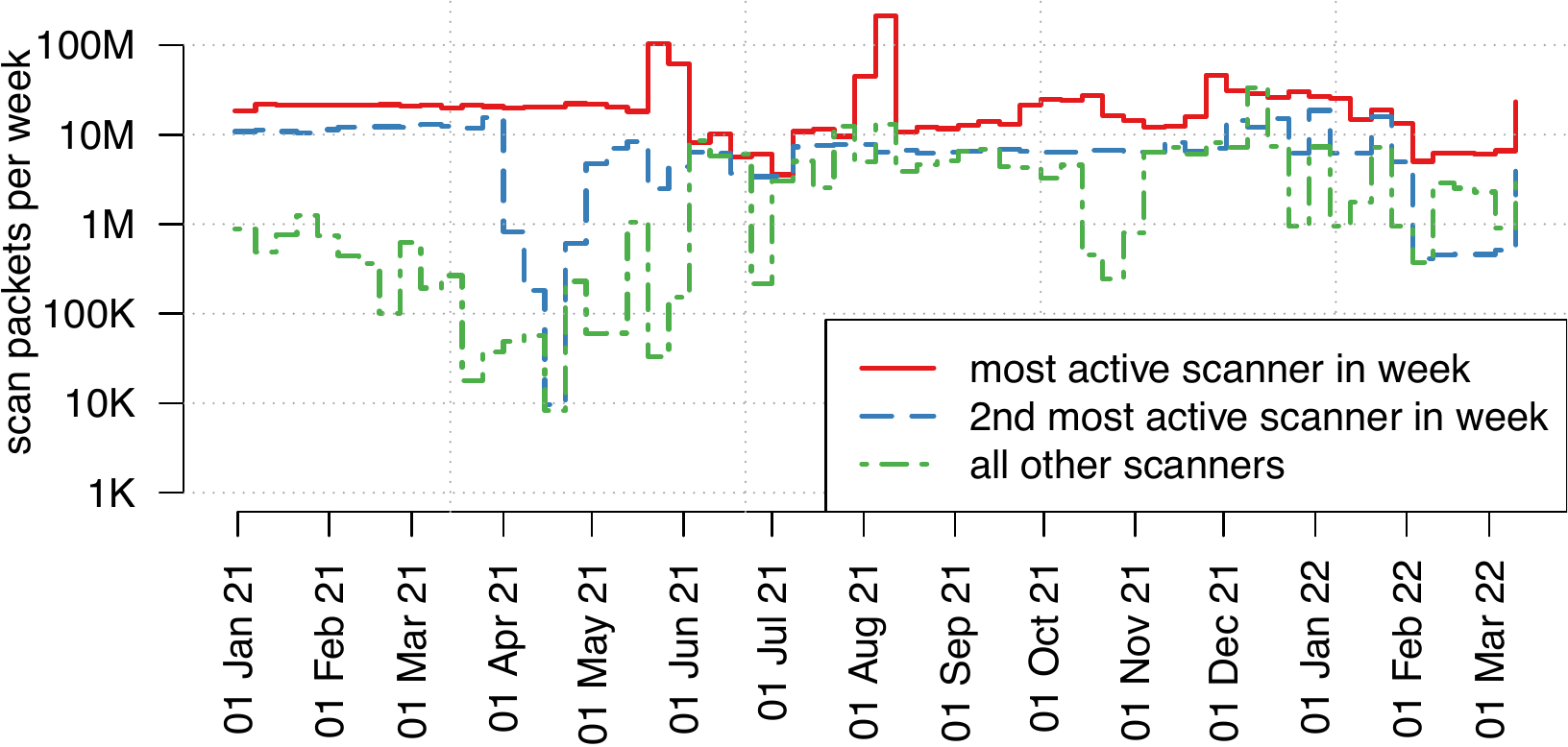}
	\caption{Weekly scan packets (/64 source aggregation).}
	\label{fig:scanpkts}
\end{figure}

\noindent\textbf{Scan traffic concentration:} Figure~\ref{fig:scanpkts} shows the number of weekly packets that can be attributed to scanning. Here, we show scans aggregated by /64 (other aggregations look similar for this particular metric). 
The most striking observation from Figure~\ref{fig:scanpkts} is that scan packets are heavily concentrated among the top two most active scan sources. 
In fact, the top two most active scan sources account, on average, for 92\% of scan packets on a week-by-week basis,\footnote{Note that the first and second most active weekly scanners are not the same entities throughout the entire measurement window.} and when seen across the entire measurement window, the two most active sources account for 70\% of all logged scan traffic.
While the large amount of scan traffic from the most active sources may suggest the absence of a trend and an overall consistent level of IPv6 scanning activity, we do see an increase in scan traffic from a larger number of sources in the latter half of our measurement period in early 2022 (see green dashed line closer, and sometimes exceeding, the blue and red line). When assessing IPv6 scans in future studies, the scan sources should be carefully inspected, as many (if not most) observations may again be dominated by one or two scanning actors.

\subsection{Scan Sources}
\label{sec:scan_sources}

\noindent\textbf{Top scan networks:} Table~\ref{tab:sourceases} shows the top-20 anonymized source ASes emitting scan traffic. We show both the total number of scan packets (for the /64 source aggregation) and the respective percentage.  We note that the two most heavily active scanners both originate from datacenter ASes located in China. Further inspection of the AS numbers and WHOIS records do not yield more information on the potential entity carrying out these massive, and continuous, scans of the IPv6 address space. Following these two ASes, we find a US-based cybersecurity company, 
and then a variety of US and global hosting and cloud providers.  Overall, we point out that scanning is heavily concentrated among a very small set of networks. In terms of scan traffic, the top-5 source ASes account for 92.8\% of scan packets, and the top-10 source ASes account for more than 99\% of scan packets.
Unlike in the case of IPv4, where scans originate from a large set of networks spanning various network types, scans in IPv6 are mostly limited to high-performance datacenters and cloud providers and their networks. Indeed, we do not find a single network that exclusively connects residential end-users to the Internet in our top-20 list.

\noindent\textbf{Scan source prefixes:} Table~\ref{tab:sourceases} shows the individual scan sources, i.e., /128 addresses, and when aggregating traffic by /64 prefixes, as well as /48 prefixes, prior to scan detection. This exercise shows major differences in the number of scan sources for different ASes. A key question is whether the individual scan sources, originating from the same AS, belong to the same entity, i.e., machine or institution carrying out a scan. For the topmost active scanner (first line in Table~\ref{tab:sourceases}), the scenario is clear, since all traffic is originated from a single IPv6 address. For other networks, however, we see that the number of individual /128s can easily be orders of magnitude larger than the respective number of active /64 scan sources. 

\noindent\textbf{Case studies:} In the following we show three cases. Sources in \textbf{AS~\#9}, a global transit provider, used 956 IPv6 source addresses, yet only two /64 prefixes. Closer
inspection of reverse DNS records and traceroutes reveals that both /64 prefixes are used by a well-known US security company, carrying out IPv6 scans and varying the lowest 7 - 9 bits in the source IP
addresses. Thus, in this case, all scan traffic from the entire /48 can be attributed to the same entity.  Note that this scan entity is solely responsible for the strong uptick in /128 sources in late 2021 in Figure~\ref{fig:weeklyscanners}.
\textbf{AS~\#18} shows up with relatively little traffic, but with more than 1,000 active /48 source prefixes. WHOIS records and BGP lookups reveal that all but one of these /48 prefixes belong to a /32, which is individually announced in BGP and exclusively used by a German cybersecurity company, carrying out scans. 
This scanning entity selects source IP addresses from across the entire /32 prefix. For comparison, a /32 prefix is the typical prefix size that ARIN and RIPE allocate to entire networks~\cite{arinv6allocations}. %
When we apply our definition of a scan to the aggregate /32, for this particular prefix, we detect 1.9 million scan packets, more than three times the number of packets we detected from this entire AS when
aggregating to /48 prefixes, as there were other /48's within the /32 that individually did not receive sufficient probes to meet our definition of a scan.  Likewise, the table reports more /48's than /64's.
Thus even /48 prefixes were insufficient to detect and classify the activity of this scanner in its entirety. 
\textbf{AS~\#6}, a global cloud provider hands out very specific IPv6 prefixes (more specific than /96s) to its customers and VMs, and we see scanning activity from 205 individual source addresses in this AS. Yet, these individual sources aggregate up to just 15 individual /64s.  The allocation policy of this particular cloud provider serves as a critical example of why using a fixed, and coarse aggregation mask for scan detection comes with the risk of aggregating different sources and entities together, and, in operational settings, possibly cause collateral damage when scan detection results in blocklisting.
More details on AS \#6 are in~\Cref{sec:heavy-hitters}.

\begin{table}
	\resizebox{.97\columnwidth}{!}{
		\begin{tabular}{r|l|r|r|r|r}
			
			\multicolumn{3}{c}{} & \multicolumn{3}{c}{\textbf{scan sources}} \\
			
			\textbf{rank} & \textbf{AS type} & \textbf{packets} & \textbf{/48s} & \textbf{/64s} & \textbf{/128s} \\
			\hline

		\#1 & Datacenter (CN) & 839M (39.2\%) & 1 & 1 & 1\\
		\#2 & Datacenter (CN) & 744M (34.8\%) & 1 & 1 & 5\\
		\#3 & Cybersecurity (US) & 275M (12.9\%) & 1 & 1 & 12\\
		\#4 & Cloud (US/global) & 78M (3.7\%) & 2 & 2 & 512\\
		\#5 & Cloud (DE) & 48M (2.3\%) & 3 & 59 & 59\\
		\#6 & Cloud (US/global) & 45M (2.1\%) & 10 & 15 & 205\\
		\#7 & Cloud (US/global) & 39M (1.8\%) & 9 & 9 & 123\\
		\#8 & Cloud (CN) & 30M (1.4\%) & 5 & 5 & 53\\
		\#9 & Transit (global) & 11M (0.5\%) & 1 & 2 & 956\\
		\#10 & Cloud (CN) & 10M (0.5\%) & 1 & 1 & 7\\
		\#11 & Cloud (US/global) & 4.7M (0.2\%) & 1 & 1 & 353\\
		\#12 & Datacenter (CN) & 3.1M (0.1\%) & 9 & 12 & 19\\
		\#13 & ISP (VN) & 2.5M (0.1\%) & 1 & 1 & 1\\
		\#14 & Datacenter (CN) & 1.6M ($\leq$ 0.1\%) & 1 & 1 & 2\\
		\#15 & Research (DE) & 1.1M ($\leq$ 0.1\%) & 1 & 1 & 1\\
		\#16 & ISP (RU) & 0.9M ($\leq$ 0.1\%) & 1 & 1 & 2\\
		\#17 & University (DE) & 0.8M ($\leq$ 0.1\%) & 1 & 1 & 2\\
		\#18 & Cloud/Transit (DE) & 0.6M ($\leq$ 0.1\%) & 1,092 & 1,057 & 1,057\\
		\#19 & ISP (RU) & 0.6M ($\leq$ 0.1\%) & 1 & 1 & 1\\
		\#20 & University (DE) & 0.5M ($\leq$ 0.1\%) & 1 & 1 & 1\\
			
			\hline
		\end{tabular}
	}
	\caption{Top 20 source ASes by scan packets over the entire measurement window (packets shown for /64 source aggregation). The number of /48 scan sources can exceed /64s or /128s if the combined traffic from the /48 satisfies the scan definition, but traffic subsets from more specific prefixes do not (e.g., in the case of AS \#18).} %
	\label{tab:sourceases}
\end{table}

\subsection{Targeting: Ports and Addresses}
\label{targeting_ports_and_addresses}

In the following, we discuss the ports and addresses targeted by scans. We report data for /64 source aggregation and separately report on AS~\#18 of Table \ref{tab:sourceases} as it contains 80\% of /64's, and would obscure attributes of the remaining /64's.

\begin{figure}
 	\centering
 	\includegraphics[width=0.8\linewidth]{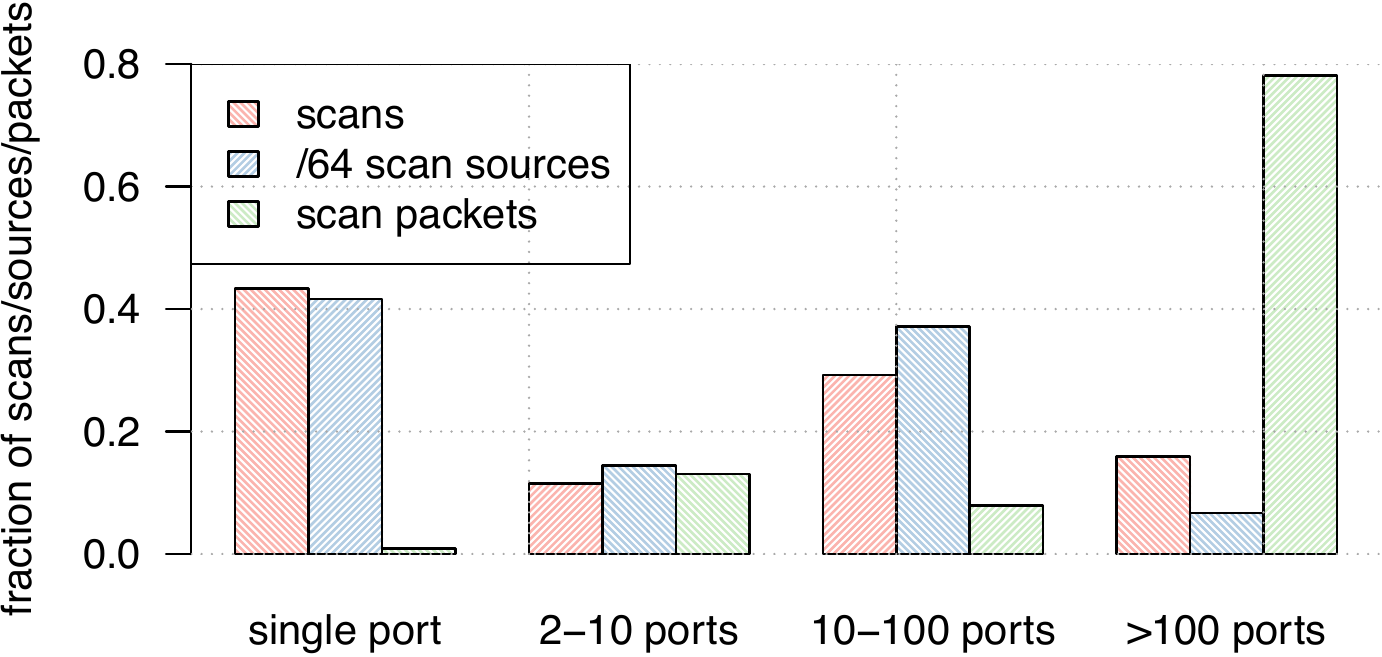}
 	\caption{Breakdown of scans, sources, and scan packets by the number of ports targeted in a scan. Scan traffic and scan sources are dominated by scanners that target multiple ports on our machines.}
 	\label{fig:port_multiport}
 \end{figure}

\noindent\textbf{Targeted ports:} We are interested in studying which potentially vulnerable services scanners target. In a first step, we study whether scanners tend to target a single port number, or
multiple port numbers. Figure~\ref{fig:port_multiport} shows the fraction of scans, scan sources, and scan packets, partitioned by how many ports an individual scan targets. 
Surprisingly, we find that the majority of detected scans and scan sources target multiple port numbers, and some of the most active scans target more than 100 ports on our machines. In fact, close to 80\% of all logged scan packets are part of heavily active scanners targeting more than 100 port numbers.  We find that the scan source in AS~\#1 targets some 444 different ports in continuous scans throughout the first half of 2021, and then changes strategy and only TCP ports 22, 3389, 8080, and 8443 are seen starting in May 2021.  The scan source in AS~\#2 targets a set of $\approx$ 635 port numbers, while AS~\#3 targets almost the entire TCP port space (45K ports). We show that this observation holds for different source aggregation levels in~\Cref{sec:multiport_agg_check}.

Table~\ref{tab:topports} shows the top-10 port numbers\footnote{Recall that port TCP/80 and TCP/443 are not in the dataset, Section~\ref{sec:vantage_point_and_dataset}.} targeted by fraction of packets, scans, and source /64s.  Given that many scans and scan sources target a broad range of port numbers, in contrast to IPv4, we do not find a clear-cut set of heavily targeted port numbers. We rather speculate that many scanners
target a large swath of well-known port numbers without focusing on exploiting a specific one.\footnote{AS \#18 of Table \ref{tab:sourceases} is not among this group as it probes just port TCP/22.} 
In the IPv4 space, in contrast, scanners targeting a single vulnerability are more typical, fueled mainly by scans carried out by botnets~\cite{richter2019scanning}.

\noindent\textbf{Targeted addresses:} 
\label{sec:targeted_addresses}
Since the IPv6 address space is so vast, a natural question is: how did the scanner determine the addresses to probe?
We begin to investigate this question by considering whether the probed address exists in DNS as the result of a forward lookup, or not.
We consider a portion of the telescope consisting of 160,000 address pairs, where one address is in DNS, and the other is not, and where the two are close in address space, often within a /123. 
For each /64 scan source, we determine the number of distinct probed addresses that are in DNS, and the number that are not. 
We find that for 75\% of the /64's \textit{all} of the probed addresses are in DNS,
while at the other end, for 10\%  of the /64's at least 33\% of the probed addresses are \textit{not} in DNS.
For AS \#18 of Table~\ref{tab:sourceases}, 50\% of the probed IPs are not in DNS. When considering the number of distinct addresses probed, we find that the larger scans tend to be the ones that have a higher fraction of not-in-DNS targets.

Regarding how a source might discover the not-in-DNS targets, a reasonable scenario is a target is found via DNS and then the scanner probes other addresses that are nearby. We consider this scenario with a sample of the /64 sources for which the target IPs are at least 50\% not-in-DNS. 
For each source, and for each not-in-DNS target we note the condition whether there was a previous and ``nearby'' in-DNS probe, where ``nearby'' means being in the same IPv6 /124, /120, /116, or /112.
Regardless of the various measures of nearby, we get mixed results.  Excluding the strictest sense of nearby of /124, one source had the nice result that the condition pertained for \textit{all} of the not-in-DNS targets.  For two other sources, this was true for 97\% of the not-in-DNS targets; for other sources, it pertained for only about half.

Considering all of the scans, while the majority of targeted addresses at the CDN are likely exposed via DNS, our data shows that some targets are discovered by other means.

\begin{table}
	\resizebox{1\columnwidth}{!}{
		
		\begin{tabular}{r|lr|lr|lr}
			\textbf{rank} & \multicolumn{2}{c|}{\textbf{total pkts}} & \multicolumn{2}{c|}{\textbf{total scans}} & \multicolumn{2}{c}{\textbf{total /64s}} \\
			\hline
			\#1 & TCP/22 & 3.5\%& TCP/22 & 45.3\%& TCP/1433 & 59.5\%\\
			\#2 & TCP/3389 & 3.4\%& TCP/23 & 43.6\%& TCP/22 & 44.2\%\\
			\#3 & TCP/8443 & 3.4\%& TCP/8080 & 42.3\%& TCP/23 & 43.9\%\\
			\#4 & TCP/8080 & 3.3\%& TCP/25 & 39.4\%& TCP/21 & 43.1\%\\
			\#5 & TCP/23 & 0.4\%& TCP/8443 & 38.3\%& TCP/8080 & 42.8\%\\
			\#6 & TCP/25 & 0.4\%& TCP/3389 & 38.3\%& TCP/3389 & 39.8\%\\
			\#7 & TCP/21 & 0.3\%& TCP/21 & 37.1\%& TCP/8000 & 39.4\%\\
			\#8 & TCP/110 & 0.2\%& TCP/5900 & 37.1\%& TCP/3128 & 39.4\%\\
			\#9 & TCP/995 & 0.2\%& TCP/993 & 36.3\%& TCP/110 & 39.0\%\\
			\#10 & TCP/8888 & 0.2\%& TCP/8081 & 36.2\%& TCP/8443 & 38.7\%\\
			\hline
		\end{tabular}

	}
	\caption{Top ports targeted by fraction of scan packets, scan events, and /64 scan sources. Since many scans target multiple port numbers, we do not find a clear-cut set of most heavily scanned services. Recall that TCP/80 and TCP/443 are not captured by our firewall.}	
	\label{tab:topports}
\vspace{-1em}
\end{table}

\section{Cross-check with Public Data}
\label{sec:mawi-crosscheck}

To increase reproducibility and generalizability of our work, we cross-check some of our salient findings with what can be observed in publicly available data.

\noindent\textbf{MAWI dataset scan detection:}  We use the MAWI dataset, which includes traffic captured at a transit link of the Japanese WIDE working group to an upstream ISP~\cite{mawi}. MAWI provides 15 minute traffic captures per day, and we leverage all available data for the same measurement window (all of 2021 until March 2022). We ought to highlight that the MAWI dataset is fundamentally different than the CDN dataset in terms of available time range (15 minutes per day), traffic volume, client networks, and geographical location. Therefore, we refrain from exact side-by-side comparisons and instead check for similar patterns where appropriate.
For scan detection, we leverage an extended version of previous work's scan definition \cite{fukuda2018knocks}, i.e., we detect a scan if a source (\textit{i}) targets at least 100 destination IPs (as opposed to 5~\cite{fukuda2018knocks}) (\textit{ii}) all packets target the same destination port, (\textit{iii}) the source sends fewer than 10 packets on the same port per destination IP, and (\textit{iv}) the entropy of the packet length is smaller than 0.1. 
In a second step, we aggregate together scans executed by a given source that targeted different ports.
In the following, we present data based on /64, and in~\Cref{sec:mawi-scanning-agg-filter} we present other aggregation levels as well as a comparison to previous work's 5 destination IPs threshold.

\noindent\textbf{Common scan activity:}
With the above scan definition,
we find a median number of six scan sources per day. Consistent with the CDN observations, both with a strict (100 destination IPs) as well as with a more loose definition of scans (5 destination IPs), we see a relatively
stable number of scans and a heavy concentration among a small number of highly active sources.
Indeed, the single most active scan source dominates almost all observed days, contributing 92.8\% of all scanning packets throughout the measurement period.
We obtained non-anonymized MAWI snapshots for selected days, and we can confirm that the most active scan source 
in the MAWI traces is the same as the most active one seen from the CDN (AS \#1, a Chinese datacenter). This underlines the, apparently, massive bandwidth available to this scan source, which has been persistently visible in both vantage points, and is still actively scanning as of May 2022. We also confirm the exact same port targeting of this scanner (hundreds of ports in early 2021, now only 6 TCP ports, \ie 22, 80, 443, 3389, 8080, and 8443). Furthermore, we find the same source IP address reported more than 2,000 times on \texttt{abuseipdb.com}, an open-source address reputation list~\cite{abuseipdb}.

\noindent\textbf{ICMPv6 scans:}\footnote{Recall that our CDN dataset does not include ICMPv6 probes, see Section~\ref{sec:vantage_point_and_dataset}.} 
In the MAWI traces, 
we find large-scale ICMPv6 scans occurring on 342 out of 439 measurement days, and on 236 of these days, ICMPv6 scan sources comprise the majority of scan sources. 
We also observe two massive peaks: On July 6, 2021 we see the first large peak in ICMPv6 scanning packets. The top scan source consists of 7 source IPs from the same /124 prefix, sending ICMPv6 echo
requests from a cybersecurity AS (AS \#3 in \Cref{tab:sourceases}). 
This specific scanning event was also noticed by other network operators and discussed on the NANOG mailing list \cite{nanogsixma}. 
The by-far largest scan peak was on December 24, 2021 and can be attributed to a single /128 belonging to a US cloud provider (not in \Cref{tab:sourceases}). 
The scan has a staggering rate of 214 kpps visible at the MAWI vantage point, suggesting that the Internet-wide scanning rate is potentially even higher.

\noindent\textbf{Address targeting:} 
To gain some insight into target addresses seen in MAWI, we compute the Hamming weight of the Interface ID (IID, the lowest 64 bits) of the target addresses of the highly active Chinese scanner, AS~\#1, and the US-based scanner in AS~\#3 of Table~\ref{tab:sourceases}. We find that for both scanners, the IIDs have a low Hamming weight and hence are not randomly generated. We point out that both sources do target DNS-exposed, as well as some non-DNS-exposed IP addresses of the CDN, suggesting that they employ other means than \textit{just} DNS to find target addresses. We also investigate target closeness for the scanners in AS \#1 and AS \#3. For both scanners we find that they target far apart addresses, resulting in a median of only 2 targeted addresses per /64 destination prefix. 
However, we also find other targeting patterns: Every packet of the ICMPv6 scanner on December 24, for example, targets a distinct /64 destination prefix, and the Hamming weights of target address IIDs follow a normal distribution, suggesting random IID generation.
We provide more details on address targeting in~\Cref{sec:mawi-scanning-agg-filter} and leave further investigation of how scanners generate target addresses, and how they may have learned non-DNS targets, for future work.

\section{Discussion}
\label{sec:discussion}

\parax{Scanning in IPv4 and IPv6:} Our findings show that large-scale scans of the IPv6 space are still comparably rare, carried out from datacenters and clouds, in stark contrast to IPv4 scans, which are, in terms of scan sources, often dominated by botnets~\cite{richter2019scanning,mirai-botnet-usenix-security}. A key aspect contributing to these differences is likely that \textit{scanning IPv6 is hard}.  Given the vastness of the space and the futility of purely random scans, there is, as of today, no way to effectively carry out IPv6 scans, e.g., from low-powered IoT devices performing random scans of the address space, like in the IPv4 space. 
Another key difference 
is that many scanners target a broad range of ports, a pattern that is rather typical for general penetration testing, as opposed to, e.g., botnets that target and exploit a specific vulnerability, often for lateral spreading~\cite{mirai-botnet-usenix-security}. 
It may well be that the relative rarity of large-scale IPv6 scans is simply the result of the inability to ``cheaply'' find destination addresses to probe. However, we argue this situation may quickly change if and when targetable IPv6 addresses become more available, be it due to advances in target generation algorithms, or exposure of addresses, e.g., via peer-to-peer applications or other rendezvous mechanisms employed by future applications. Future work involves continuous monitoring and re-appraisal of the evolving landscape of IPv6 scanning, as well as more detailed assessment of how today's scanners find target IP addresses.

\parax{Scan detection and attribution:} A key challenge we faced throughout this work is the required level of source address aggregation to properly isolate an individual scanning entity. While some of the most active scan sources use a single IPv6 address,
at the other extreme, we found a scanning actor that used an entire /32 routed prefix. 
Note that a /32 is also the default allocation for \textit{an entire network} as per ARIN and RIPE.
Inferring a source prefix that is too specific can miss portions of the scanning activity, and too coarse can
merge traffic from multiple scanning actors and/or non-scanning hosts, e.g. in the case of cloud providers. If, in operational settings, scan detection leads to blocklisting, too coarse aggregation may lead to collateral damage, i.e., blocking of traffic from legitimate sources.
Intrusion Detection Systems (IDSes) should determine the aggregation in real-time.
One idea is to dynamically change aggregation levels, e.g., start on a non-aggregated level and then adjust to larger prefixes.
Another idea is to track simultaneously various aggregations.
IDSes may have to rely on traffic features and other header fields to fingerprint individual scans and hosts.

\parax{Related work:} The research community has developed and investigated strategies to find IPv6 target addresses for scanning~\cite{gasser2018clusters, murdock2017target, foremski2016entropy,cui20206gcvae,yang20226graph,song2022det}. The matter of variable prefix lengths and allocations in the IPv6 space has been studied in several related works~\cite{padmanabhan2020dynamips,kIP,akamai-v6-imc15}. Scanning activity in the IPv4 space was studied from multiple vantage points, see~\cite{richter2019scanning, internetwidescan, darknet-imc15} and references therein. Passive detection of IPv6 scanning activity has received comparably less attention, in part due to a lack of suitable vantage points. Fukuda and Heidemann propose to leverage DNS backscatter to detect IPv6 scans~\cite{fukuda2018knocks} and Tanveer et al. propose to attract potential IPv6 probing and scans by actively sending out probing traffic~\cite{tanveer2022glowing}. 

To the best of our knowledge, our work presents the first longitudinal study on large-scale IPv6 scanning activity, as seen from firewall logs captured at the edge of a major Content Distribution Network.

\subsection*{Acknowledgments} We thank the anonymous reviewers and our shepherd Amogh Dhamdhere for the valuable feedback, and Kenjiro Cho for providing access to the MAWI packet traces.

\bibliographystyle{ACM-Reference-Format}
\bibliography{paper}

\appendix

\section{Appendix}

\subsection{CDN Filtering Artifacts}
\label{sec:artifacts}

A prime example of such artifacts are unsuccessful \texttt{SMTP} attempts, where an SMTP server tries to deliver email to an address associated with a domain hosted by the CDN. If there is no
\texttt{MX} record configured for the respective domain, SMTP falls back to \texttt{A} and \texttt{AAAA} records. In our logs, such events manifest as source IP addresses repeatedly trying to access
a specific service on some, or sometimes many\footnote{The mapping process of the CDN maps repeated connection requests of a client to a potentially large number of CDN machines.} IP addresses of the
telescope, hence showing characteristics of a scan (single source targeting many destinations). Another common example are hosts trying to establish IPsec connections, sending ISAKMP packets on UDP
port 500. Similarly to the SMTP case, if a host gets mapped to a significantly large number of CDN machines over time, such attempts show the properties of potential scanning behavior.\footnote{We note that such artifacts are more noticeable in IPv6, compared to IPv4, likely because of they are obscured by the high volume of random scans targeting the IPv4 space.}

Looking at the dominant protocols in traffic filtered out (November 2021), we find that \texttt{UDP/500} (ISKAMP/IPsec) and \texttt{TCP/25} (IMAP) are the two most prevalent protocols by packets and source addresses.

\subsection{Further Analysis of MAWI Trace}
\label{sec:mawi-scanning-agg-filter}

\begin{figure}
	\centering
	\includegraphics[width=.9\linewidth]{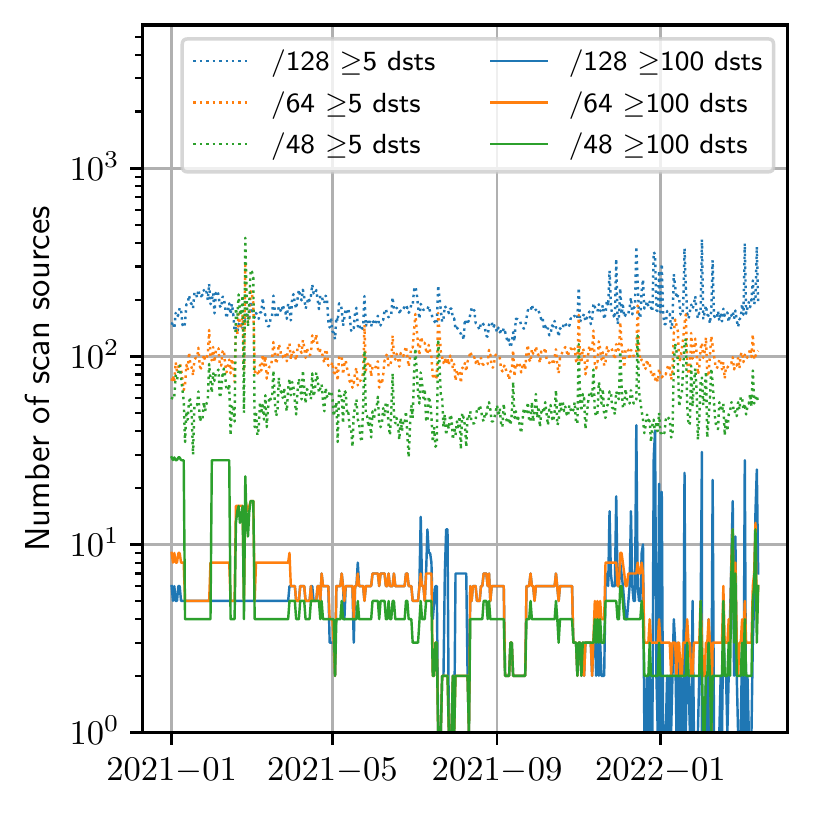}
    \caption{MAWI: Number of scan sources for different aggregations and destination IP filters over time.}
    \label{fig:mawi-scanning-agg-filter}
\end{figure}

\begin{figure}
	\centering
	\includegraphics[width=.9\linewidth]{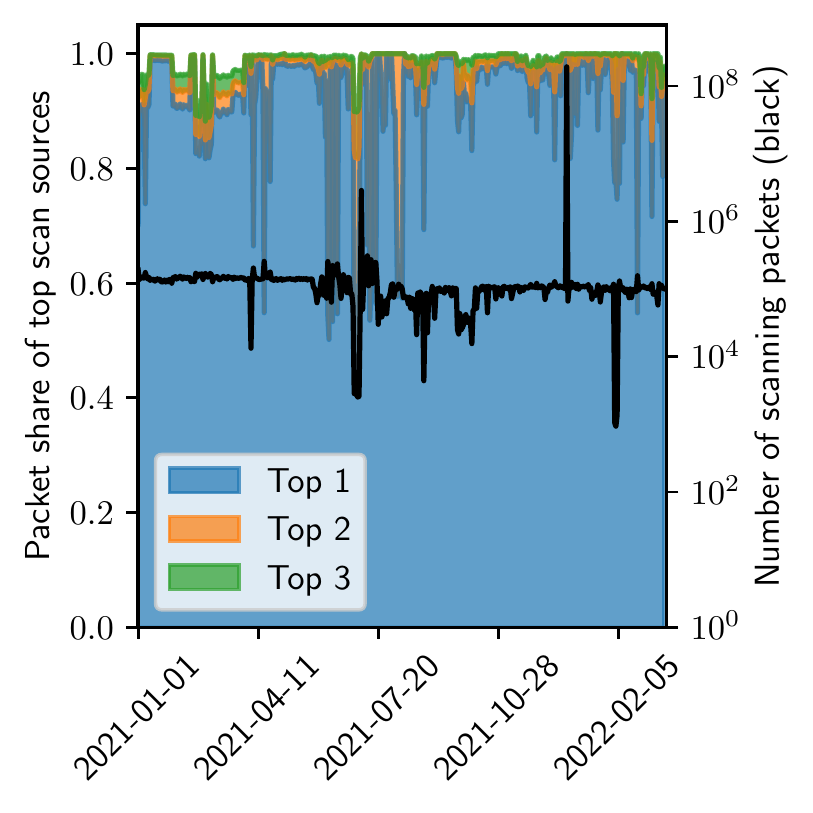}
	\caption{MAWI: Packet share sent by top 1, top 2, and top 3 scan sources over time (left y-axis); number of scanning packets (right y-axis).}
	\label{fig:mawi-top-scanners-packets}
\end{figure}

\parax{Scan detection and traffic share:}
Figure~\ref{fig:mawi-scanning-agg-filter} shows the number of daily scan sources when applying aggregation per /48, /64, or /128 and requiring a minimum number of either 5 (for comparability with Fukuda and Heidemann \cite{fukuda2018knocks}) or 100 (our large-scale scan definition) destination IP addresses. 
For all three aggregations and both definitions, Figure~\ref{fig:mawi-scanning-agg-filter} shows a relatively constant number of scan sources over the 15 months,
which is consistent with the CDN data, as shown in Figure~\ref{fig:weeklyscanners}.
Also note that with the threshold of 5, one infers more than an order of magnitude more scan sources.
Figure~\ref{fig:mawi-top-scanners-packets} shows the number of daily scan packets (see \textit{y}-axis on the right side of the plot), and the share of the three most active scan sources (see \textit{y}-axis on the left side of the plot). Similarly to the CDN observations, we find that scan traffic is heavily concentrated among the top most active scan sources.

\parax{Target address MAWI comparison:} Next, for the selected days where we have non-anonymized MAWI data, we investigate whether the targets are chosen from sources such as the IPv6 hitlist \cite{gasser2018clusters}.
We find that the Chinese scan source (AS \#1 in \Cref{tab:sourceases}) has an almost non-existent overlap with the IPv6 hitlist.
One exception is May 27, 2021, where we find an astonishing 99.2\% overlap with addresses also present in the IPv6 hitlist, although with much fewer unique destination addresses (dropping from 50k+ to just 2.3k).
We see a similar behavior for this scanner in the CDN dataset, although not as pronounced as in MAWI.
Note that this is also the day, where this source switched from targeting hundreds of ports to just six TCP ports (cf. \Cref{sec:targeted_addresses}).
We also investigate the two peaks on July 6 and December 24 (cf. \Cref{sec:mawi-crosscheck}) and find no overlap with the IPv6 hitlist.

\begin{figure}
	\centering
	\includegraphics[width=0.8\linewidth]{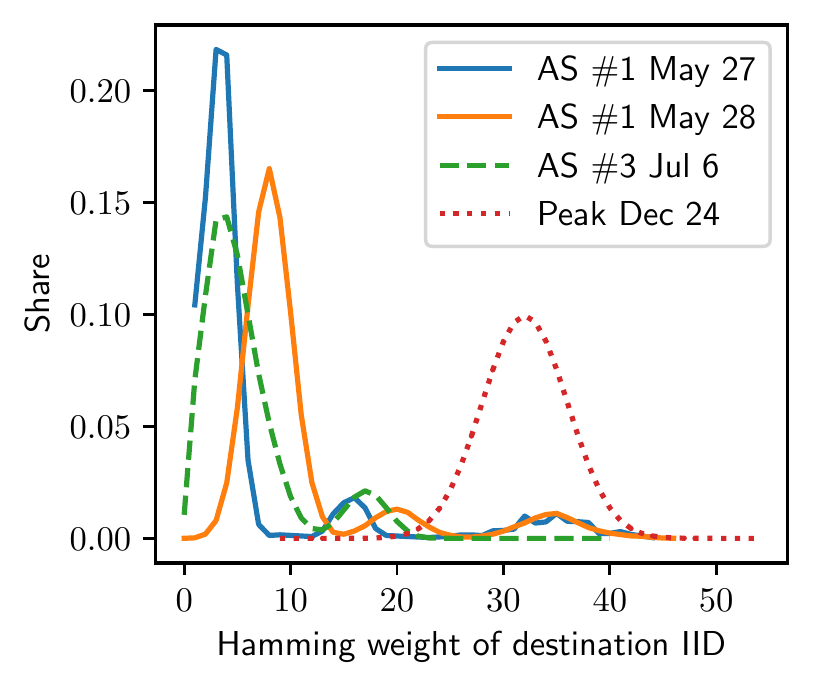}
	\caption{MAWI: Hamming weight distribution for select scan sources and dates.}
	\label{fig:mawi-hamming-weight}
\end{figure}

To better understand the target selection technique of scan sources not relying on the IPv6 hitlist, we analyze the randomness of targeted addresses.
Similar as previous work \cite{gasser2016scanning}, we use the Hamming weight (\ie the number of bits set to '1') of the rightmost 64 bits (\ie the IID) as an indicator for destination address randomness.
\Cref{fig:mawi-hamming-weight} shows the Hamming weight distribution for selected scan sources and dates.
We can see that the destinations of the Chinese scan source (AS \#1) has a relatively low Hamming weight (HW).
Interestingly, we see that May 27 has an even lower HW compared to May 28.
This is due to the fact that on May 27 this source almost exclusively targeted addresses found in the IPv6 hitlist, which exhibit a lower entropy.
This source switched from probing already known addresses (likely to generate a seed set of active addresses) to discovering new IPv6 addresses \cite{murdock2017target,liu20196tree,cui20206gcvae,cui20206veclm,cui20216gan,hou20216hit,yang20226graph,song2022det}.
The MAWI scan peak on July 6 caused by AS \#3 has a similarly low HW, indicating that this source is also looking for unknown active addresses instead of focusing on known ones.
An outlier in terms of HW distribution is the peak on December 24, 2021, originating from a US cloud provider.
The HW distribution follows a perfect Gaussian distribution, suggesting that this source used completely randomizes destination addresses instead of relying on prior knowledge or structural properties \cite{foremski2016entropy} of the IPv6 address space.

\subsection{Ports per Scan at /128 and /48 Aggregations}
\label{sec:multiport_agg_check}

\begin{figure}
	\centering
	\begin{subfigure}{0.98\linewidth}
		\includegraphics[width=0.9\linewidth]{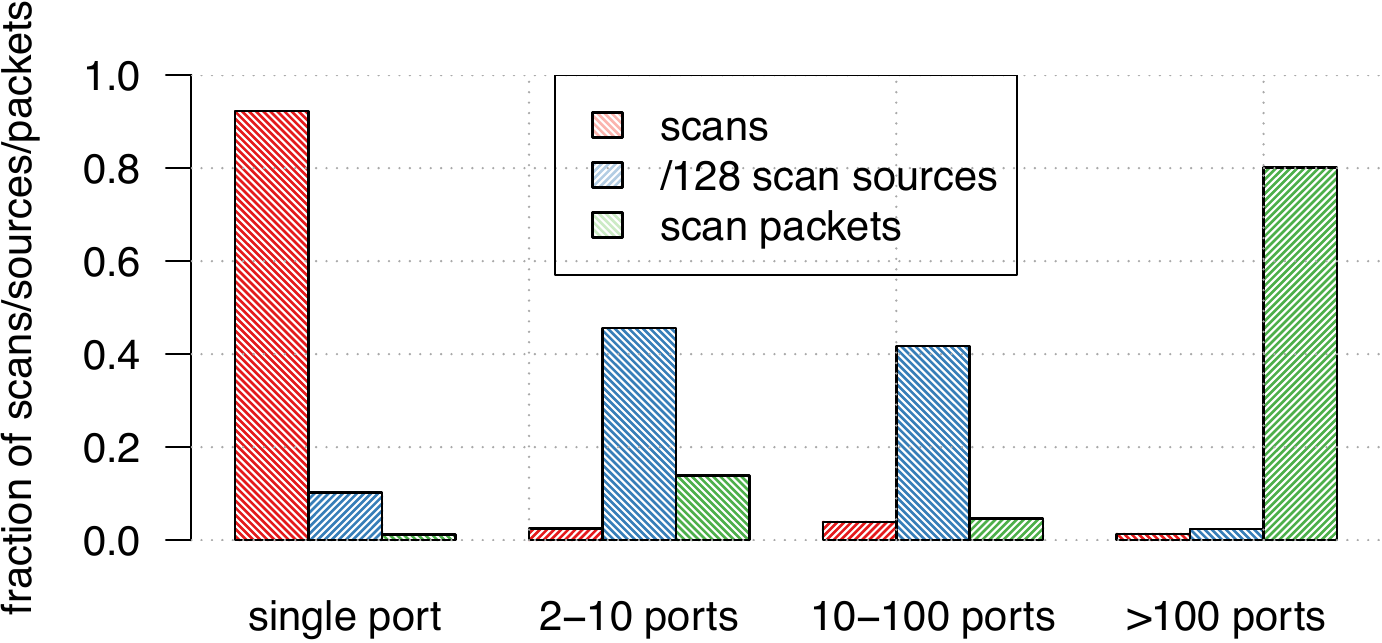}
		\caption{Number of ports per scan without aggregation (/128 sources). }
		\label{fig:multiport128}
	\end{subfigure}
	\hspace{0.1em}
	\begin{subfigure}{0.98\linewidth}
		\includegraphics[width=0.9\linewidth]{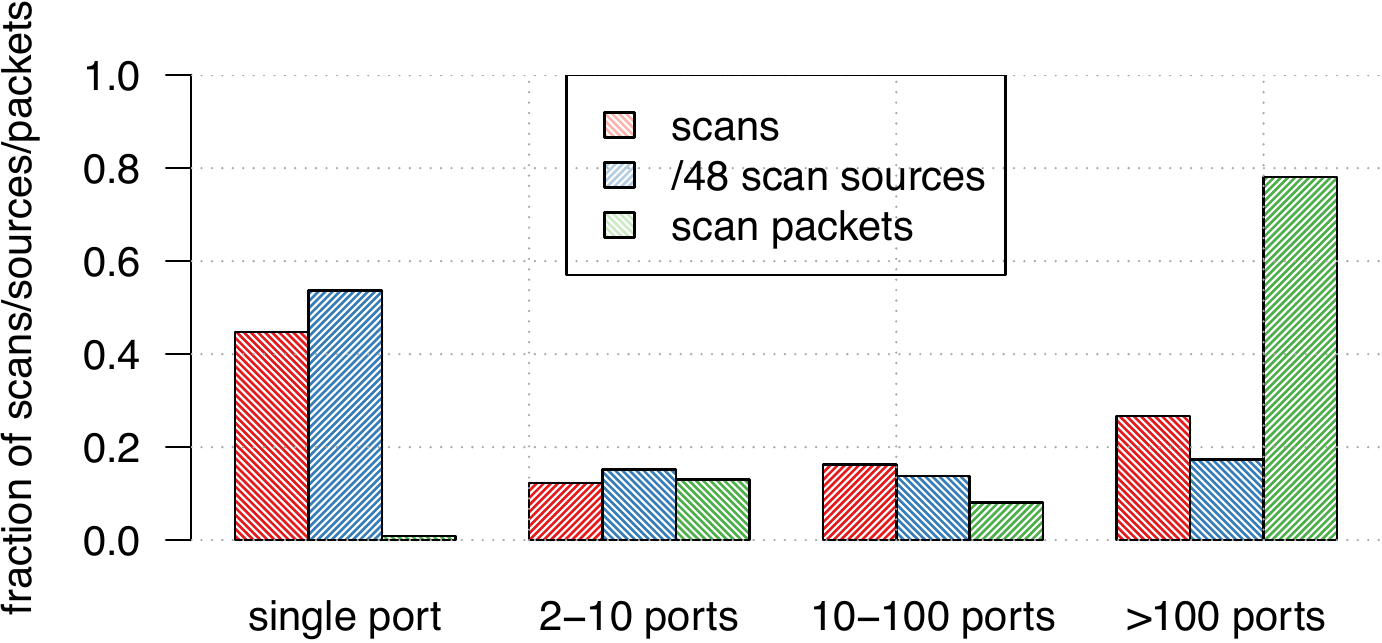}
		\caption{Number of ports per scan with /48 aggregation.}
		\label{fig:multiport48}
	\end{subfigure}
	\caption{Ports targeted per scan for /128 (no aggregation) and/48 (heavy source aggregation).}
	\label{fig:multiport_different_aggregations}
\end{figure}

Figure~\ref{fig:multiport_different_aggregations} shows the number of ports targeted in scans when not aggregating scan sources, i.e., treating each /128 source individually, and when aggregating scan sources to entire /48 prefixes. In either aggregation the statement that most packets can be attributed to multi-port scans holds.\footnote{For each scan entity, we count the packets for each port number and get the fraction \textit{f} of packets that hit the most common port in that scan. If \textit{f} is larger than 0.5 we tag the scan as single port, else if \textit{f} is larger than 0.09 we tag the scan to target less than 10 port numbers, else if \textit{f} is larger than 0.009 we tag the scan as targeting less than 100 port numbers, and more than 100 port numbers otherwise. This way, we will not misclassify a scan to be multi-port if only a tiny fraction of its corresponding packets target a myriad of port numbers.} Treating scans without aggregation, the number of single-port scans (but not sources) increases dramatically, caused by one scanning entity that scans for different port numbers progressively in distinct scanning episodes. The heavy /48 aggregation shows a larger fraction of sources contributing to scans targeting more than 100 port numbers which may be result of aggregating different scanning entities together.

\subsection{Cloud Provider \#6 of Table \ref{tab:sourceases} }
\label{sec:heavy-hitters}

Although unlikely, it is possible that a single actor is responsible for all of the observed activity.
However, different origin locations or probed targets and ports or activity periods and intensity do not necessarily distinguish between actors using a public cloud.
Though, as a partial result, very similar activity across addresses is suggestive of common control.
For example, leveraging the analysis of ``Targeted Addresses'' of Section \ref{targeting_ports_and_addresses},
two of the fifteen /64's have very similar values for number of addresses probed that are in-DNS, 71386, 71354, respectively, and not-in-DNS, 63483, 64547, and the fraction in-DNS is the same to three
significant figures.
Both had the same number of scans.
Both had hits at the start and end of the 15 month observation interval, indicating the scanning started before and continued thereafter.
The targeted addresses largely coincided - the intersection divided by union was 78\%.
Both probed all ports a various number of times, though one did three times as many probes.
Collectively, these similarities strongly suggest a common actor.
As an aside, the two /64's are in different /48's and thus are an example of an scanning actor using prefixes from separate address space.

\end{document}